\documentclass[iop,apj]{emulateapj}
\usepackage{natbib}
\newcommand{\threefifty}{\ensuremath{350\:\mu\mathrm{m}}}
\newcommand{\paperone}{Paper I}
\newcommand{\als}{ALS03}

\slugcomment{Accepted in Astrophysical Journal}
\begin{document}

\author{Nicholas L.\ Chapman\altaffilmark{1},
Jacqueline A.\ Davidson\altaffilmark{2},
Paul F.\ Goldsmith\altaffilmark{3},
Martin Houde\altaffilmark{4,5},
Woojin Kwon\altaffilmark{6,7},
Zhi-Yun Li\altaffilmark{8},
Leslie W.\ Looney\altaffilmark{6},
Brenda Matthews\altaffilmark{9,10},
Tristan G.\ Matthews\altaffilmark{1},
Giles Novak\altaffilmark{1},
Ruisheng Peng\altaffilmark{11},
John E.\ Vaillancourt\altaffilmark{12},
Nikolaus H.\ Volgenau\altaffilmark{13}}

\altaffiltext{1}{Center for Interdisciplinary Exploration and Research in
Astrophysics (CIERA) \& Dept.\ of Physics \& Astronomy, Northwestern University,
2145 Sheridan Road, Evanston, IL 60208}

\altaffiltext{2}{School of Physics, University of Western Australia,
35 Stirling Hwy, Crawley, WA 6009, Australia}

\altaffiltext{3}{Jet Propulsion Laboratory, California Institute of Technology,
4800 Oak Grove Drive, MS 264-782, Pasadena, CA 91109}

\altaffiltext{4}{Department of Physics and Astronomy, University of Western
Ontario, London, ON, Canada}

\altaffiltext{5}{Division of Physics, Mathematics and Astronomy, California
Institute of Technology, Pasadena, CA 91125}

\altaffiltext{6}{Department of Astronomy, University of Illinois, 1002 West
Green Street, Urbana, IL 61801}

\altaffiltext{7}{SRON Netherlands Institute for Space Research, Landleven 12,
9747 AD, Groningen, The Netherlands}

\altaffiltext{8}{Astronomy Department, University of Virginia, Charlottesville,
VA 22904}

\altaffiltext{9}{Herzberg Institute of Astrophysics, National Research Council
of Canada, 5071 West Saanich Road, Victoria, BC V9E 2E7, Canada}

\altaffiltext{10}{Department of Physics and Astronomy, University of Victoria,
3800 Finnerty Road, Victoria, BC V8P 1A1, Canada}

\altaffiltext{11}{Caltech Submillimeter Observatory, 111 Nowelo Street, Hilo, HI
96720}

\altaffiltext{12}{SOFIA Science Center, Universities Space Research Association,
NASA Ames Research Center, MS 232-11, Moffett Field, CA 94035-0001}

\altaffiltext{13}{California Institute of Technology, Owens Valley Radio
Observatory, Big Pine, CA 93513}

\title{Alignment Between Flattened Protostellar Infall Envelopes and
Ambient Magnetic Fields}

\doublespace

\begin{abstract}

We present \threefifty{} polarization observations of four low-mass cores 
containing Class 0 protostars: L483, L1157, L1448-IRS2, and Serp-FIR1.  This is 
the second paper in a larger survey aimed at testing magnetically regulated 
models for core-collapse. One key prediction of these models is that the mean 
magnetic field in a core should be aligned with the symmetry axis (minor axis) 
of the flattened YSO inner envelope (aka pseudodisk).  Furthermore, the field 
should exhibit a pinched or hour-glass shaped morphology as gravity drags the 
field inward towards the central protostar.  We combine our results for the 
four cores with results for three similar cores that were published in the 
first paper from our survey.  An analysis of the \threefifty{} polarization 
data for the seven cores yields evidence of a positive correlation between mean 
field direction and pseudodisk symmetry axis.  Our rough estimate 
for the probability of obtaining by pure chance a correlation as strong as the 
one we found is about 5\%.  In addition, we combine together data for multiple 
cores to create a source-averaged magnetic field map having improved 
signal-to-noise ratio, and this map shows good agreement between mean field 
direction and pseudodisk axis (they are within $15^\circ$).  We also see hints 
of a magnetic pinch in the source-averaged map.  We conclude that core-scale 
magnetic fields appear to be strong enough to guide gas infall, as predicted by 
the magnetically regulated models.  Finally, we find evidence of a positive 
correlation between core magnetic field direction and bipolar outflow axis.

\end{abstract}

\section{Introduction}\label{sec:introduction}

Star formation occurs in molecular clouds. Even though these clouds are
relatively large, with sizes $\sim$few parsecs, they are gravitationally bound.
Despite being bound, the inferred rate of star formation in the clouds is less
than that estimated from gravitational free-fall collapse. To explain this
discrepancy, a mechanism is needed to regulate the star formation rate.
Two such mechanisms that have been proposed are magnetic support and
turbulence. See, e.g., \citet{mckee07} for a review of these issues.
This turbulence is often modeled as being super-Alfv{\'e}nic, though
trans-Alfv{\'e}nic turbulence may also act to suppress star formation
\citep{federrath13}.

Embedded within molecular clouds are individual cores (typically $\sim$$20,000$
AU in size for low-mass cores), each of which could potentially collapse to form
a star or group of stars. Besides their possible role in supporting clouds
against gravity, magnetic fields may also strongly affect the evolution of these
individual cores.  Magnetically regulated core-collapse models
\citep{shu87,galli93a,galli93b,tomisaka98,allen03b,allen03a,shu04} treat star
formation in low-mass cores where magnetic fields are dynamically important
compared to turbulence. These models have two key features which can be
observationally tested.  First, there should exist a flattened inner core having
a size of a few thousand AU, known as a pseudodisk. This pseudodisk is not a
rotationally supported disk, but instead is formed by the preferential collapse
of material along magnetic field lines as opposed to across field lines.  Thus,
the core magnetic field should be parallel to the symmetry (minor) axis of the
pseudodisk. Secondly, the magnetic field inside the infall radius should have a
pinched hourglass morphology.  Outside the infall radius, the magnetic field
should be uniform or only gently pinched.

A third observational test is to compare the axis of the bipolar outflow with
the core magnetic field direction.  Outflows are believed to be launched from
the inner regions of Keplerian circumstellar disks, with the outflow axes
aligned with the rotation axes of the disks \citep[e.g.,][]{konigl00,shu00}.  If
such disks form with their rotation axes aligned with the core-scale field, then
one might expect a positive correlation between outflow axes and core magnetic
fields. However, the theoretical expectation regarding the alignment of
circumstellar disk axes with core fields is unsettled at this time.  On the one
hand, since magnetic braking proceeds faster for angular momentum components
perpendicular to a near-uniform, large-scale field during cloud core 
contraction before collapse \citep[e.g.,][]{mouschovias79}, one might expect
that the rotation axis of the core and circumstellar disk should align with the
core magnetic field.  This point of view is advocated in \S\,4.3 of
\citet{allen03b}.  However, it is not currently understood how a rotationally
supported disk can form at all in this scenario, since magnetic braking in a 
collapsing core should remove angular momentum sufficiently rapidly to prevent
disk formation \citep[e.g.,][]{mellon08}.  Recent simulations of magnetized core
collapse by \citet{joos12} suggest that circumstellar disks can form more easily
if the core magnetic field and core rotation axis are not aligned.  In such a
scenario, it seems unlikely that disks would form with their axes preferentially
parallel to the core magnetic field.

The direction of the plane of the sky component of the magnetic field can be
observed through polarization measurements.  Interstellar dust grains align with
their long axes preferentially perpendicular to the local magnetic field.
Polarization arises because the dust grains preferentially absorb and emit
radiation with the electric field vector parallel to the grain's long axis.  At
optical and near-infrared wavelengths, one observes polarized absorption of
background starlight seen through a cloud.  Thus, the inferred magnetic field is
parallel to the polarization direction.  At submillimeter and millimeter
wavelengths, the dust grains emit polarized radiation and the inferred magnetic
field is orthogonal to the polarization direction.  See \citet{lazarian07} for a
review of the theory of magnetic dust grain alignment.

Because the outflow is larger and more easily observed than the pseudodisk, most
tests for magnetic regulation of core-collapse have focused on measuring
alignment between outflow axes and magnetic field directions.  For example,
\citet{menard04} found no correlation when they compared the axes of T-Tauri
star outflows in the Taurus molecular cloud with magnetic field directions
inferred from optical and near-IR polarimetry.  Similarly, \citet{targon11} used
protostars with a range of ages and also found no correlation between outflow
axis and field direction.  One possible conclusion from these studies is that
circumstellar disks form with their axes oriented randomly with respect to the
fields in the natal cores.  An alternative possibility is that outflows from
T-Tauri stars will have injected more turbulence into their surroundings, in
comparison with outflows from the much younger Class 0 sources.  This additional
turbulence could scramble any initial alignment between the disk/outflow system
and the core magnetic field.  Indeed, when \citet{targon11} focused on just the
Class 0 and Class I sources in their sample, they did find a statistical
alignment between outflow axis and magnetic field direction.  Observations of
polarized dust emission at $850\:\mu$m in Bok globules have found the magnetic
field to be nearly aligned with the outflow axis or pseudodisk symmetry axis in
some sources, but in other sources the field is closer to being perpendicular to
the outflow axis \citep[e.g.,][]{henning01,vallee03,wolf03}. High resolution
interferometric submillimeter polarimetry of NGC 1333 IRAS 4A has revealed the
first clear case of a pinch morphology in a low-mass star-forming core
\citep{girart06}.  The symmetry axis of the pinched magnetic field was found to
be nearly aligned with the minor axis of the inner core. Subsequently,
\citet{attard09} mapped the magnetic field in this core on larger spatial scales
and found a fairly uniform field running parallel to the symmetry axis of the
small-scale hourglass field.

As illustrated by these examples, previous work has found some cases of
alignment between magnetic fields and outflows or magnetic fields and
pseudodisks, but a number of counter examples as well.  However, the variety of
evolutionary stages, stellar masses, stellar multiplicities, and spatial scales
probed in these studies confuses comparisons with  core-collapse models.  To
address these shortcomings, we began a submillimeter polarimetric survey of
low-mass, isolated (i.e.\ single), nearby ($\lesssim400$ pc), young (Class 0)
YSOs with well-defined bipolar outflows. Furthermore, we attempted to include
only YSOs whose outflows lie nearly parallel to the plane of the sky, although
we were not always successful in this regard, as discussed later in this
section.  Our source selection criteria aim to ensure that the objects in our
sample have simple geometries that are not confused by nearby objects and are
close enough that we can resolve small size scales ($\leq 4000$ AU) in the
centers of cores. Thus, our survey is providing data for direct comparison with
magnetically regulated core-collapse models.  The first paper from this survey
presented results for B335, L1527, and IC348-SMM2 \citep[hereafter
\paperone]{davidson11}.  These three cores exhibit flattened central regions
consistent with their being edge-on pseudodisks.  The symmetry axis of the
pseudodisk in each core is nearly parallel to the outflow axis.  The magnetic
fields in cores show some degree of agreement with the predictions of
magnetically regulated core-collapse models, but \paperone{} concluded that more
data were needed for definitive tests.

\begin{deluxetable*}{llrrc}
\tablewidth{0pt}
\tablecaption{\label{tab:obs} SHARP \threefifty{} Polarimetry Observations}
\tablehead{\colhead{Source} & \colhead{Dates Observed} & \colhead{No.\ HWP} & 
\colhead{$\tau_{\threefifty}$\tablenotemark{a}} & 
\colhead{Chop Throw\tablenotemark{b}}\\
\colhead{} & \colhead{} & \colhead{Cycles} & \colhead{} & 
\colhead{(arcsec)}}
\startdata
L483       & 2009 Sept 20-22    &  34 & 1.3-1.8 & 300 \\
           & 2010 Mar 27        &  13 & 0.8-0.9 & 300 \\
           & 2010 June 2-3      &  11 & 0.8-1.3 & 180,300 \\
\hline
L1157      & 2008 Sept 6-10     & 136 & 0.6-1.5 & 300 \\
\hline
L1448-IRS2 & 2009 Sept 17,20-22 &  92 & 1.2-1.9 & 300 \\
           & 2009 Nov  8        &  21 & 1.4-1.7 & 180 \\
\hline
Serp-FIR1  & 2009 Sept 21       &   5 & 1.3     & 300 \\
           & 2010 June 4        &  11 & 1.8     & 300
\enddata
\tablenotetext{a}{Zenith atmospheric opacity at \threefifty}
\tablenotetext{b}{Chop distance in cross-elevation}
\end{deluxetable*}

In the present paper we expand the survey by presenting results for four new
sources: L483, L1157, L1448-IRS2, and Serp-FIR1. In \S\,\ref {sec:observations}
we discuss the observations and the data reduction, and we present our
measurements. In \S\,\ref{sec:maps} we show the inferred magnetic field maps for
each of the four cores, and we provide information (compiled from the
literature) concerning bipolar outflows and pseudodisk-like structures in the
cores. Despite our best efforts in choosing sources, recent work suggests that
Serp-FIR1 has its outflow aligned nearly parallel to the line-of-sight
\citep{enoch09b}. For the other three cores, just as for the cores in \paperone,
the outflows are much less inclined with respect to the plane of the sky
(estimated inclination angles never exceed $40^\circ$). In \S\,\ref{sec:seven}
we present a combined analysis of our current survey sample of seven cores,
including four from the present paper and three from \paperone.  We test for the
predicted alignment between core magnetic field direction and pseudodisk
symmetry axis, and we also test for a correlation between core magnetic field
direction and outflow axis. In addition, to increase the signal-to-noise ratio
we create a source-averaged magnetic field map by combining measurements from
multiple sources.  These analyses make use of previously published estimates of
outflow inclination angle for each of our seven cores. In
\S\,\ref{sec:discussion} we discuss the implications of our results for
understanding magnetic effects in star formation. Lastly, in
\S\,\ref{sec:conclusions} we summarize our results.

\section{Observations and Results}\label{sec:observations}

Polarimetric observations of L483, L1157, L1448-IRS2, and Serp-FIR1 were
obtained using the SHARP polarimeter during five observing runs at the Caltech
Submillimeter Observatory (CSO).  The runs took place during the period
September 2008--June 2010. SHARP \citep{li08} is a fore-optics module that adds
polarimetric capabilities to SHARC-II, a $12\times32$ pixel bolometer array used
at the CSO \citep{dowell03}.  SHARP separates the incident radiation into two
orthogonal polarization states that are then imaged side-by-side on the SHARC-II
array.  SHARP includes a half-wave plate located upstream from the polarizing
splitting optics.  The wavelength of observation was \threefifty{} and the
effective beam size was $\sim$$10\arcsec$.  Polarimetric observations with SHARP
involve carrying out chop-nod photometry at each of four half-wave plate
rotation angles; a single such cycle requires approximately seven minutes.
Additional details concerning the observations are listed in Table
\ref{tab:obs}.

A full discussion of our data acquisition and reduction procedures is given in
\paperone. Here, however, we will go into some detail on the calculation of
errors since this was done differently in the present paper than in \paperone.
The Stokes parameters $I$, $Q$, and $U$ represent the total ($I$) and linearly
polarized ($Q$, $U$) flux and are derived from the flux detected at each of the
four HWP angles during a single cycle. For the analysis presented in \paperone,
the authors divided the data into subsets and computed the reduced chi-squared,
$\chi^2_r$, among them to check if the results were consistent within the
nominal uncertainties. The $\chi^2_r$ values from that analysis ranged from 1.5
to 2.1, suggesting that the nominal errors were too small. The ``extra errors''
are caused by correlated noise between pixels (covariance). The errors were
shown to occur on relatively short time scales, so they were treated as if they
were statistical in nature, i.e.\ the nominal errors were inflated by the square
root of $\chi^2_r$ to produce the final errors.

For the present paper we handled these extra errors by using the generalized
Gauss-Markov theorem, which statistically accounts for the covariance between
pixels \citep[e.g.,][]{cox06}.  In addition to computing the variance of each
pixel in the Stokes parameter maps for a given cycle, we also calculated the
covariance between each pair of pixels.  When the single-cycle Stokes parameter
maps for a core were combined to create the final Stokes parameter maps, the
generalized Gauss-Markov theorem was used to propagate the variances and
covariances and compute the final uncertainties for each pixel.  By including
the covariances in this way, the resulting uncertainties became larger. However,
the reduced chi-squared values computed during consistency checking became
smaller.  Specifically, our $\chi^2_r$ values were 1.04 (L483), 1.19 (L1157),
0.99 (L1448-IRS2), and 1.21 (Serp-FIR1).  Since these reduced chi-squared values
were near unity, there was no reason to inflate our nominal errors.  The net
result of our new covariance analysis method is that the final signal-to-noise
is mostly unaffected compared to the older chi-squared inflation method.
However, the processing is more straightforward.

Our combined Stokes parameter maps contain polarization measurements for every
$9\farcs5$ grid point.  The grid is centered at the position of the source's
peak flux. These Stokes parameters  were converted to percentage polarization
($p$) and polarization angle  ($\theta$). Since polarization cannot be negative,
this leads to a small positive polarization bias, for which we corrected
(\citealp{hildebrand00}; \citealp[see also][]{vaillancourt06}).  We considered
any polarization measurement having $p/\sigma_p \geq 2$ after debiasing to be a
detection.  Our polarization detections are listed in Table \ref{tab:pol}.  Note
that our cutoff is at $2\sigma$, rather than a more traditional (and
conservative) $3\sigma$ threshold.  With a $2\sigma$ threshold, the
uncertainties in polarization angle range up to almost $13^\circ$.  However,
because we are only testing the gross predictions of magnetically regulated
core-collapse models, this cutoff level is acceptable.

\begin{deluxetable}{lrrrrrrr}
\tablewidth{0pt}
\tablecaption{\label{tab:pol}SHARP $350\:\mu$m Polarimetry Results}

\tablehead{\colhead{Source} & \colhead{$\Delta\alpha$\tablenotemark{a}} & 
\colhead{$\Delta\delta$\tablenotemark{a}} & \colhead{$p$} & 
\colhead{$\sigma_p$} & \colhead{$\theta$\tablenotemark{b}} & 
\colhead{$\sigma_\theta$\tablenotemark{b}} & \colhead{I\tablenotemark{c}}\\
\colhead{} & \colhead{($''$)} & \colhead{($''$)} & \colhead{(\%)} &
\colhead{(\%)} & \colhead{(deg)} & \colhead{(deg)} & \colhead{(\%)}}

\startdata
L483 &  19.0 & -28.5 & 17.4 &  7.9 &  27.6 & 10.2 &  14 \\
     &  19.0 & -19.0 &  9.3 &  4.1 &   5.7 & 10.9 &  17 \\
     &   9.5 & -19.0 &  5.7 &  2.2 &  12.8 & 10.1 &  20 \\
     &   9.5 &   0.0 &  1.0 &  0.5 &  -7.4 & 12.7 &  78 \\
     &   0.0 &  19.0 &  4.6 &  1.5 &  41.6 &  8.6 &  29 \\
     &  -9.5 &   0.0 &  0.8 &  0.4 &   5.7 & 12.4 &  71 \\
     & -19.0 & -19.0 &  3.5 &  1.3 &  -1.8 &  9.9 &  27 \\
     & -19.0 &  -9.5 &  1.7 &  0.7 & -13.5 & 11.6 &  41 \\
     & -28.5 & -19.0 &  6.0 &  2.3 & -23.2 & 10.0 &  20 \\
     & -38.0 &   0.0 &  7.5 &  3.5 & -19.2 & 11.7 &  17 \\
\hline
L1157 &  19.0 & -19.0 &  7.4 &  3.5 & -85.2 & 10.4 &  11 \\
      &   0.0 &   0.0 &  0.7 &  0.2 &  52.1 &  9.0 & 100 \\
      &  -9.5 &  19.0 &  5.8 &  2.5 &  40.1 & 10.6 &  11 \\
      & -19.0 &   9.5 &  5.4 &  2.6 &  41.5 & 12.0 &  11 \\
\hline
L1448 IRS2 &  38.0 &  -9.5 &  9.5 &  3.2 &  61.2 &  9.0 &  31 \\
          &  38.0 &   0.0 &  9.5 &  3.2 &  34.8 &  8.7 &  29 \\
          &  28.5 & -19.0 &  9.6 &  3.3 &  63.5 &  8.9 &  23 \\
          &  19.0 &   0.0 &  4.0 &  1.5 &  50.2 & 10.1 &  31 \\
          &   9.5 & -38.0 & 14.5 &  5.8 &  74.4 &  9.7 &  20 \\
          &   9.5 & -28.5 &  7.5 &  2.7 &  75.1 &  9.6 &  25 \\
          &   9.5 &   0.0 &  2.2 &  0.8 &  50.4 &  9.7 &  53 \\
          &   9.5 &   9.5 &  2.5 &  1.2 &  51.4 & 12.2 &  38 \\
          &   9.5 &  19.0 &  7.6 &  2.9 &  56.1 & 10.1 &  19 \\
          &  -9.5 &  19.0 &  9.6 &  3.9 &  73.8 & 10.4 &  16 \\
          & -19.0 & -28.5 &  7.5 &  3.5 &  74.1 & 12.2 &  21 \\
          & -19.0 &  -9.5 &  4.9 &  2.3 &  55.9 & 12.0 &  23 \\
          & -19.0 &   9.5 &  7.3 &  3.4 &  65.4 & 11.9 &  18 \\
\hline
Serp FIR1 &  19.0 &  -9.5 &  5.1 &  1.7 &   7.1 &  8.2 &  18 \\
          &   9.5 & -19.0 &  3.5 &  1.4 &   2.8 & 10.1 &  19 \\
          &   0.0 & -38.0 & 14.8 &  7.1 &  12.6 &  9.3 &   8 \\
          &   0.0 &   9.5 &  1.0 &  0.4 & -22.7 & 10.0 &  46 \\
          &  -9.5 &  -9.5 &  1.6 &  0.8 & -30.1 & 12.0 &  25 \\
          &  -9.5 &   0.0 &  1.2 &  0.4 & -31.2 &  8.9 &  41 \\
          & -19.0 & -19.0 &  3.7 &  1.7 &  -2.2 & 11.4 &  13 \\
          & -19.0 &  -9.5 &  3.3 &  1.2 & -44.2 &  9.1 &  15 \\
          & -19.0 &   0.0 &  2.3 &  1.1 & -45.2 & 12.4 &  15 \\
          & -28.5 & -19.0 &  6.9 &  3.2 &  14.5 & 10.8 &   9
\enddata

\tablenotetext{a}{Offsets in arcseconds from positions listed in Table \ref{tab:prop}}
\tablenotetext{b}{Position angle of the polarization $E$-vector, measured
east of north}
\tablenotetext{c}{Intensity as a percentage of the peak for each core}

\end{deluxetable}

\section{Polarization Maps for Individual Sources}\label{sec:maps}

\begin{deluxetable*}{lr@{\hskip 18pt}r@{\hskip 18pt}r@{\hskip 18pt}r}
\tablewidth{0pt}
\tablecaption{\label{tab:prop}Basic Source Properties}
\tablehead{\colhead{Information} & \colhead{L483} & \colhead{L1157} & 
\colhead{L1448-IRS2} & \colhead{Serp-FIR1}}
\startdata
Right Ascension (J2000)            &  18 17 29.8\tablenotemark{(1)} &  20 39 06.2\tablenotemark{(6)} & 03 25 22.3\tablenotemark{(6)}  & 18 29 49.6\tablenotemark{(16)} \\
Declination (J2000)                & -04 39 38.3\tablenotemark{(1)} &  68 02 15.9\tablenotemark{(6)} & 30 45 13.3\tablenotemark{(6)}  & 01 15 21.9\tablenotemark{(16)} \\
Distance (pc)                      &  $200\pm30$\tablenotemark{(2)} &  $325\pm13$\tablenotemark{(7)} & $232\pm18$\tablenotemark{(12)} &  $415\pm5$\tablenotemark{(17)} \\
Pseudo-disk P.A. (deg)\tablenotemark{a} &    36\tablenotemark{(3)}  &          75\tablenotemark{(8)} &     45\tablenotemark{{(6,13)}} & \nodata \\
Infall Radius (AU)                 &        8000\tablenotemark{(4)} &        8500\tablenotemark{(9)} &       8000\tablenotemark{(14)} &       5000\tablenotemark{(18)} \\
\cutinhead{Outflow Properties}
Position Angle (deg)\tablenotemark{a}    &       105\tablenotemark{(5)} &   155\tablenotemark{(10)}  &       138\tablenotemark{(15)}  & 130\tablenotemark{(19)} \\
Inclination Angle (deg)\tablenotemark{b} & $40\pm10$\tablenotemark{(5)} &     9\tablenotemark{(11)}  & $33\pm^{8}_6$\tablenotemark{(14)} & $72.5\pm7.5$\tablenotemark{(18)}
\enddata
\tablenotetext{a}{Position angles denote the angle of the long axis of the
pseudodisk or outflow, measured east of north.}
\tablenotetext{b}{The inclination angle is measured with respect to the plane of
the sky.}
\tablerefs{(1) \citet{jorgensen04}; (2) \citet{prato08}; (3) \citet{fuller00};
(4) \citet{myers95};    (5) \citet{fuller95}; (6) \citet{kwon09};
(7) \citet{straizys92}; (8) \citet{looney07}; (9) \citet{gueth97};
(10) \citet{davis95};  (11) \citet{gueth96}; (12) \citet{hirota11};
(13) \citet{chen10};   (14) \citet{tobin07}; (15) \citet{davis08};
(16) \citet{harvey07}; (17) \citet{dzib10};  (18) \citet{enoch09b};
(19) \citet{rodriguez89}}
\end{deluxetable*}

In this section we will compare our submillimeter polarization data with the
observed outflow direction, pseudodisk position angle, and the predictions of
the magnetically regulated core-collapse model of \citet[hereafter
\als]{allen03b}, just as was done in \paperone.  The \als{} model numerically
computes the gravitational collapse of a singular isothermal core that is
magnetized and rotating with its rotation axis aligned with the large-scale
magnetic field. To compare our SHARP data with this model, we need the
pseudodisk position angle and core infall radius for each source. Using the
literature, we compiled these and other relevant properties for each source and
we list them in Table \ref{tab:prop}.  We give a detailed discussion of how the
information in Table \ref{tab:prop} was obtained in \S\,\ref{sec:l483} through
\S\,\ref{sec:serp}.

\subsection{Overview of Polarization Maps}\label{sec:overview}

The red and blue lines (vectors) in Figures \ref{fig:sharpL483} through
\ref{fig:sharpSerp} show the inferred magnetic field directions for each core in
a manner similar to Figures 2 through 7 in \paperone{}. These inferred field
directions are orthogonal to the measured directions of the \threefifty{}
polarization. In the left panel, the vectors are plotted with lengths
proportional to the percentage polarization.  As in \paperone, we distinguish
between high-flux and low-flux regions as a way to flag polarization
measurements that may have large contamination from the parent cloud. Red
vectors are used for sight-lines where the flux is greater than or equal to
$25\%$ of the peak flux, while blue vectors indicate sight-lines not meeting
this threshold.  The $25\%$ flux cutoff matches the level used in \paperone.

The left panel of each figure shows in grayscale a \emph{Spitzer} $4.5\:\mu$m
waveband image.  This waveband is an excellent tracer of outflows because it
contains a shocked H$_2$ emission line.  For L483 and Serp-FIR1, images are from
the final delivery of the \emph{Cores to Disks} Legacy Science
program\footnote{http://irsa.ipac.caltech.edu/data/SPITZER/docs/\\ spitzermission/observingprograms/legacy/c2d/}
while for L1157 and L1448-IRS2 images are from the \emph{Spitzer} Heritage
Archive\footnote{http://archive.spitzer.caltech.edu/}.  Overlaid on each
\emph{Spitzer} image are contours of the observed \threefifty{} emission,
ranging from $20\%$ to $90\%$ of the peak flux in steps of $10\%$.  The dashed
circle and double-headed black vector indicate the infall radius and outflow
position angle for each source, respectively (estimated from observations, see
Table \ref{tab:prop}).  The right panel plots the inferred magnetic field
vectors superposed on Figure 8(c) from \als{}. Each vector is now shown with the
same length to make the magnetic field morphology clearer.  The vector locations
are scaled to the model infall radius and the model is rotated by the pseudodisk
position angle.  It is important to note that there are no free parameters to
help fit the magnetic field vectors to the \als{} model. The infall radius and
pseudodisk position angle are set by observations.

Also shown in each figure is the mean magnetic field direction that we computed
for each core (white outlined double-headed arrow).  In \paperone{} this mean
was not computed, but we require it for the statistical analyses that we will
carry out in \S\,\ref{sec:align} below.  Because a polarization angle of
$0^\circ$ is the same as $180^\circ$, computing the mean polarization angle is
non-trivial. We use the Equal Weight Stokes Mean, defined by \citet{li06}. In
brief, this method converts each angle to Stokes $q = Q/I$ and $u = U/I$,
computes the unweighted averages $\bar{q}$ and $\bar{u}$, and then converts
$\bar{q}$ and $\bar{u}$ back to an angle.  The uncertainty in mean angle is
derived from propagation of errors.  Only the red vectors in each core are used
when computing the mean magnetic field angle.

\als{} present several models corresponding to different values of magnetic
field strength and rotation speed.  Just as in \paperone, we show only one
model, corresponding to Fig.\ 8(c) of \als, chosen because it has intermediate
values for both magnetization and rotation.  However, note that all models in
\als{} (excluding the unmagnetized one) show a similar pinch in the field.  Data
from our survey cannot yet resolve the small differences between the various
models in \als.  Also, it is important to keep in mind that \als{} display a
slice through the center of their core.  Averaging of the magnetic field along
the line-of-sight will lessen the observed pinch in the field.  Furthermore,
large inclination angles of the pseudodisk symmetry axis will distort the
observed pinch pattern.  \als{} show this distortion in their Figure 4 for
inclinations of $30^\circ$ and $90^\circ$.  At an inclination of $30^\circ$ the
distortion is mild, but at $90^\circ$ the magnetic field lines extend radially
outwards from the core center, with some twisting due to rotation.

\subsection{L483}\label{sec:l483}

L483 is located in the Aquila Rift at a distance of $200\pm30$ pc
\citep{prato08}.  The embedded protostar in L483 is IRAS 18148-0440.  Based on
its spectral energy distribution, it is usually classified as a Class 0
protostar \citep{fuller95}. However, \citet{tafalla00} argued that the outflow
from the protostar shares some properties with those observed in more evolved
sources and that IRAS 18148-0440 may thus be transitioning from Class 0 to Class
I.

The inclination angle of the outflow in L483 is not well-known.  In the present
paper we measure inclination from the plane of the sky; thus $0^\circ$
inclination means the outflow is parallel to the sky plane and $90^\circ$ means
the outflow is pointing along the line-of-sight.  \citet{fuller95} measured
differences in $2.22\:\mu$m brightness between the eastern and western lobes of
the outflow. By fitting this emission to a simple model, they estimated the
inclination angle to be $40^\circ$.  Based on their Figure 5 we estimate an
uncertainty of $\pm10^\circ$ on the inclination.  This $40^\circ$ inclination is
the only quantitative value for the inclination angle in the literature, but it
may be an overestimate.  \citet{hatchell99} measured $^{12}$CO $J=4\rightarrow3$
emission from the outflow and found some overlap between the red- and
blue-shifted emission in each outflow lobe, which suggests that the L483 outflow
has a smaller inclination angle.  We estimate the position angle of the outflow
to be $105^\circ$ based on the shocked H$_2$ emission \citep{fuller95}.

\begin{figure*}

\plotone{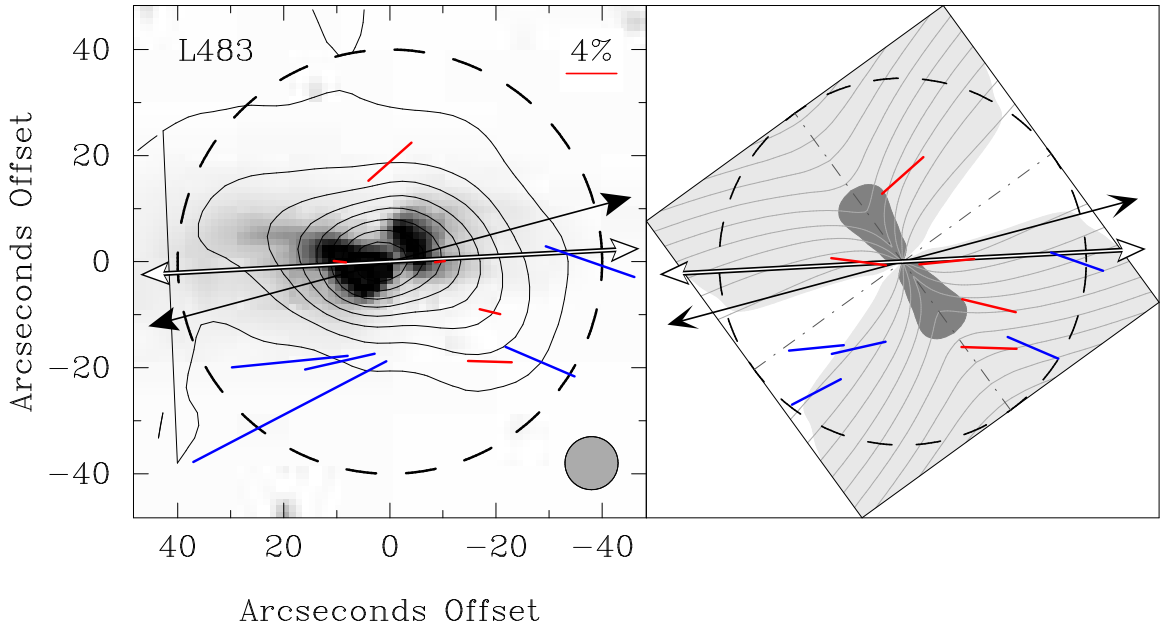}

\caption{\label{fig:sharpL483} Inferred magnetic field direction for the core
L483.  \emph{(left)} The halftone image shows the \emph{Spitzer} IRAC band 2
($4.5\:\mu$m) emission for the core.  Contours show the $350\:\mu$m intensity;
they range from 20\% to 90\% of the peak flux, in steps of 10\%.  Red and blue
vectors show the measured $350\:\mu$m polarization, where the angle of each
vector has been rotated by $90^\circ$ to show the inferred magnetic field
direction.  Vectors are plotted for points where $p/\sigma_p \geq2$ after
debiasing.  The length of each vector has been scaled in proportion to its
percentage polarization.  Red vectors indicate positions where the flux is
greater than or equal to $25\%$ of the peak flux, blue vectors indicate
positions not meeting this threshold.  The large black vector shows the outflow
direction, while the white outlined vector is the mean magnetic field direction
computed from the red vectors.  The dashed circle indicates the measured infall
radius. The gray circle at the bottom right shows the SHARP beam size
($10\arcsec$). \emph{(right)} The magnetic field vectors plotted on the
core-collapse infall model taken from Figure 8(c) of \citet{allen03b}.  All
vectors have been plotted the same length.  The dark grey region shows the model
pseudodisk.  The grey lines show the model magnetic field lines.  The
dashed-dotted lines are the orientation axes for the model pseudodisk.  The
model has been rotated to match the measured position angle of the observed
pseudodisk.}

\end{figure*}

There are no molecular line measurements of a pseudodisk in L483.  This could be
due to depletion and possibly also optical depth effects
\citep{fuller00,park00,jorgensen04}. \citet{fuller00} observed a narrow
absorption lane at $3.4\:\mu$m with a width of $\sim$$250$ AU. From visual
inspection of their Figure 4, we estimate that the diameter of this absorption
lane must be $\sim$1000--2000 AU based on the apparent $3.4\:\mu$m extinction. 
The lane can also be seen at lower resolution in the \emph{Spitzer} $4.5\:\mu$m
emission in Figure \ref{fig:sharpL483}.  It has roughly the right size to be a
pseudodisk.  We measured its position angle to be $36^\circ$, which we adopt as
the pseudodisk position angle.

There are two different estimates for the infall radius of L483. \citet{myers95}
observed N$_2$H$^+$ and C$_3$H$_2$ rotational emission lines at the source peak
and nearby positions offset from the peak.  They found that the centroid line
velocities decreased (suggesting infall motion) as positions approached the
protostar over a distance of 0.04 pc ($40\arcsec$ for a 200 pc distance).
\citet{tafalla00} detected self-absorption in the H$_2$CO ($2_{12}-1_{11}$)
emission toward L483, with a brighter blue-shifted peak compared to the
red-shifted peak, indicative of infall motions.  The radius over which they were
able to measure a stronger blue-shifted peak was $20\arcsec$. Because the
self-absorption profile requires specific physical conditions to occur, the lack
of a stronger blue-shifted peak is not necessarily evidence for lack of infall.
Therefore, we adopt an infall radius of $40\arcsec$ for this paper.

Figure \ref{fig:sharpL483} shows the inferred magnetic field in L483.  The
$20\%$ contour of the \threefifty{} emission appears distorted due to artifacts
at the edge of the map.  The magnetic field is fairly ordered with a mean
direction of $93\pm6^\circ$.  The mean field direction is offset by $12^\circ$
from the direction of the outflow and offset by $33^\circ$ from the symmetry
axis of the pseudodisk.  Considering the red and blue vectors together, there is
a suggestion of a pinch in the magnetic field.

\citet{dotson10} reported polarimetric observations of L483 at \threefifty. They
obtained several upper limits ($p + 2\sigma_p < 1\%$) for sight lines near the
flux peak and three polarization detections from the periphery (regions with $I
\lesssim 20\%$ of their peak flux).  Our two vectors closest to the flux peak
have polarization magnitudes of $1.0\%$ and $0.8\%$, consistent with the upper
limits of \citet{dotson10}.  Only one of the periphery vectors from
\citet{dotson10} overlaps regions where we detect polarization; this occurs at
$(\Delta RA, \Delta DEC) \sim(+30\arcsec,-20\arcsec)$.  We detect polarization
at two locations near this position. Taken together, and considering the coarser
angular resolution of \citet{dotson10}, our vectors agree with their measurement
in both polarization angle and polarization percentage.

\subsection{L1157}\label{sec:l1157}

The first distance estimate for L1157 was 440 pc, based on the distance to the
NGC 7023 open cluster \citep{viotti69}.  More recently, \citet{kun98} estimated
distances of 200, 300, and 450 pc for different clouds in Cepheus. Because the
Galactic latitude of L1157 is similar to that of the 200 pc and 300 pc clouds,
some recent authors have assumed a distance of 250 pc for L1157
\citep{looney07,chiang10,tobin10}.  \citet{straizys92} used an interstellar
reddening-distance relationship to estimate a distance of $325\pm13$ pc for the
L1147/L1158 subgroup (which includes L1157).  The latter distance was used by
the \emph{Spitzer} Gould's Belt Legacy Science Program \citep{kirk09} and is the
distance we adopt here.  L1157 contains the Class 0 protostar IRAS 20386+6751
\citep{gueth97}.

\begin{figure*}

\plotone{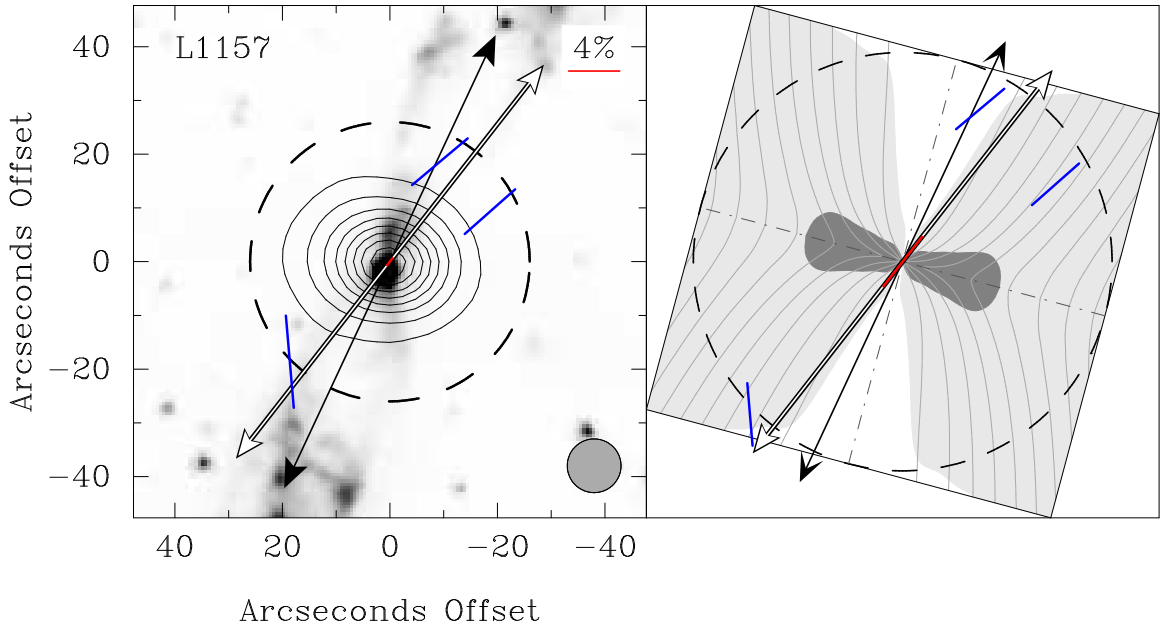}

\caption{\label{fig:sharpL1157} Same as Figure \ref{fig:sharpL483}, except
for L1157.}

\end{figure*}

The outflow in L1157 has an inclination angle of $9^\circ$ \citep{gueth96}. It
has been mapped in shocked H$_2$ by \citet{davis95} who found a position angle
of $155^\circ$.  L1157 has a flattened envelope roughly perpendicular to the
outflow.  It is very prominent in $8\:\mu$m absorption with a diameter of
$\sim$$2\arcmin$ \citep{looney07}.  \citet{chiang10} observed N$_2$H$^+$
$J=1\rightarrow0$ and found it to spatially coincide with the absorption
feature.  The full width half-maximum of the N$_2$H$^+$ is
$\sim$$11\arcsec\times18\arcsec$, or $\sim$$3600\times5900$ AU at the distance
of L1157.  The N$_2$H$^+$ emission shows evidence of both infall and rotation
\citep{chiang10} and its size is not too different from the predicted pseudodisk
size, so we adopt the position angle of the N$_2$H$^+$ emission, which is
$75^\circ$, as the pseudodisk position angle.  \citet{gueth97} observed
$^{13}$CO $J=1\rightarrow0$ and $J=2\rightarrow1$ transitions in L1157.  Both
spectra show a self-absorption profile with the blue-shifted peak stronger than
the red-shifted peak, indicating infall (see their Figure 8).  Using these two
spectra, \citet{gueth97} estimated the path length towards the central protostar
of the $^{13}$CO to be 8500 AU.  We adopt this value as the infall radius for
L1157.

We show our results for L1157 in Figure \ref{fig:sharpL1157}.  Only one of our
polarization detections has a corresponding flux value greater than or equal  to
$25\%$ of the flux peak.  This vector appears at the peak and has $>3\sigma$
significance.  It is offset by $13^\circ$ with respect to the outflow axis and
by $23^\circ$ with respect to the pseudodisk symmetry axis.  \citet{hull13}
measured 1.3 mm polarization in L1157 on spatial scales much smaller than those
studied in the present paper.  Their mean magnetic field is consistent with our
measurement, and will be discussed in more detail in \S\,\ref{sec:discussion}.

\subsection{L1448-IRS2}\label{sec:l1448}

The L1448 complex is located in the Perseus cloud.  L1448-IRS2 lies
approximately $3\arcmin$ west of L1448-mm/L1448C and L1448-IRS3 and $3\arcmin$
east of L1448-IRS1.  In addition, roughly $50\arcsec$ east of IRS2 is a
candidate first hydrostatic core labeled L1448-IRS2E \citep{chen10}.
\citet{hirota11} obtained a distance of $232\pm18$ pc towards H$_2$O masers in
L1448C. Because of L1448C's proximity, we adopt this distance for L1448-IRS2.
L1448-IRS2 contains the protostar IRAS 03222+3034.  Based on its spectral energy
distribution it is a Class 0 source \citep{olinger99}.

\begin{figure*}

\plotone{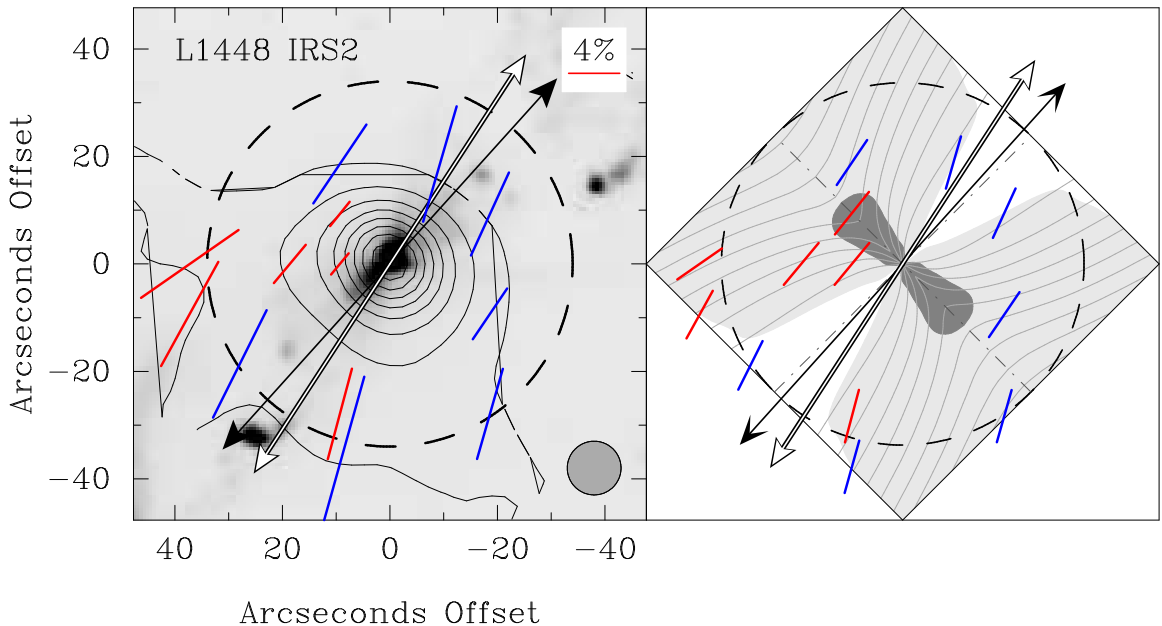}

\caption{\label{fig:sharpL1448} Same as Figure \ref{fig:sharpL483} except for
L1448-IRS2.  The two easternmost red vectors are excluded when computing
the mean field direction.  See \S\,\ref{sec:l1448} for details.}

\end{figure*}

The position angle for the outflow is $138^\circ$ as derived from shocked H$_2$
emission \citep{davis08}.  Less certain is the inclination angle of the outflow.
\citet{tobin07} used radiative transfer codes \citep{whitney03b,whitney03a} to
model continuum emission data covering the wavelength range $2.2\:\mu$m to $2.7$
mm. The best fit inclination angle is $33^{+8}_{-6}$ degrees.  A possible
pseudodisk has been observed in 1.3 mm continuum emission by both \citet{kwon09}
and \citet{chen10} with $5\arcsec$ and $3\arcsec$ resolution, respectively.  In
both maps, the core appears extended with a long axis oriented at position angle
$45^\circ$.  We measure the length of the long axis as $\sim$$14\arcsec$ from
the data of \citet{kwon09} and $\sim$$5\arcsec$ from the data of \citet{chen10}. 
Thus the spatial size is $\sim$1000--3000 AU for a distance of $232$ pc.  There
are no direct measurements of the infall radius for this source.  However, the
best-fit model from \citet{tobin07} has an envelope radius of 8000 AU.  Because
the \citet{whitney03b,whitney03a} models use an infalling envelope, we adopt the
envelope radius as the infall radius for L1448-IRS2.

\begin{figure*}

\plotone{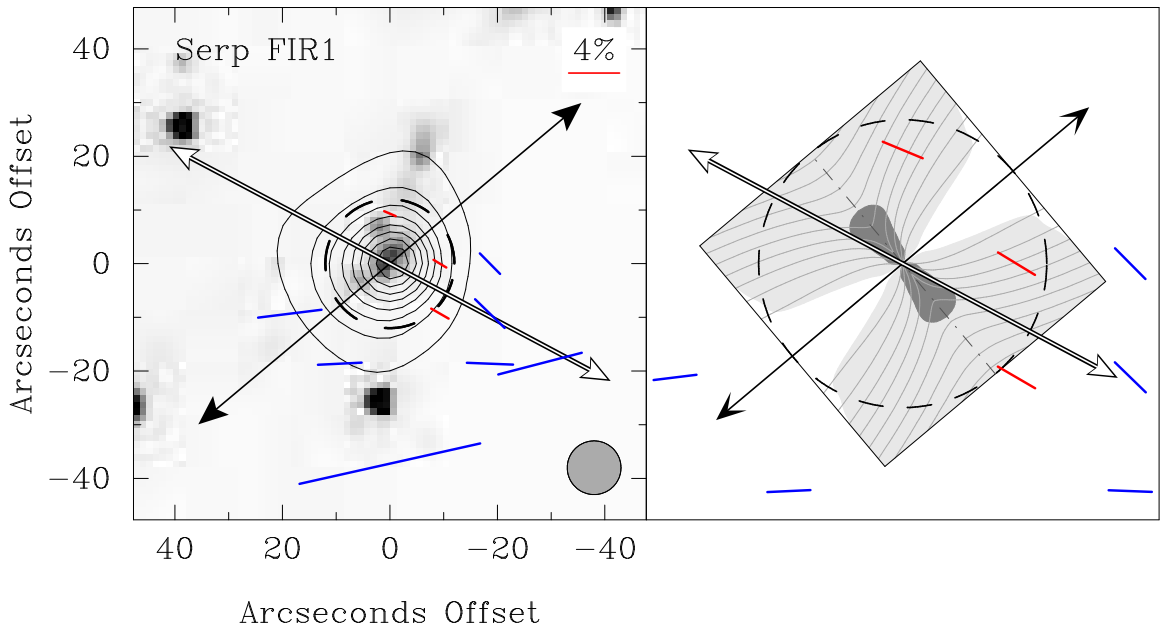}

\caption{\label{fig:sharpSerp} Same as Figure \ref{fig:sharpL483} except for
Serp-FIR1.  Because no pseudodisk has been found for this source, the model in
the right panel was rotationally aligned with the observed outflow axis.  Note
that the outflow in this source is nearly parallel to the line-of-sight.}

\end{figure*}

In Figure \ref{fig:sharpL1448} we plot our results for L1448-IRS2.  We pick up
some emission from L1448-IRS2E at the edges of our map.  This emission is the
primary reason for the distorted $20\%$ contour.  Taken together, the red and
blue vectors are very ordered with a uniform direction.  There is no evidence of
a pinch in the magnetic field.  Note that the two easternmost vectors arise from
the wings of L1448-IRS2E.  To avoid any possible bias, we exclude these vectors
and use only the remaining four red vectors in computing the mean field
direction.  The mean field direction is $147\pm5^\circ$ which is offset by
$9^\circ$ from the outflow axis and by $12^\circ$ from the pseudodisk symmetry
axis.  \citet{hull13} measured 1.3 mm polarization in L1448-IRS2 on spatial
scales much smaller than those studied in the present paper.  Their mean
magnetic field is consistent with our measurement, and will be discussed in more
detail in \S\,\ref{sec:discussion}.

\subsection{Serp-FIR1}\label{sec:serp}

Serp-FIR1 (also known as Serp-SMM1) is located in the Serpens core.  Most
distance estimates for Serpens are in the range 200--400 pc \citep{eiroa08}.  In
the present paper we adopt a distance of $415\pm5$ pc based on the work of
\citet{dzib10}, who measured the trigonometric parallax for the binary YSO EC95,
located in the Serpens cloud core.  Serp-FIR1 is the brightest source in the
Serpens core at submillimeter wavelengths and contains the Class 0 protostar
IRAS 18273+0113 \citep{hurt96}.

The Serp-FIR1 outflow has been observed in 6 cm continuum with the Very Large
Array. The continuum is resolved into three peaks, one centered on the source,
the other two offset by $\sim$$6\arcsec$ on either side \citep {rodriguez89}. 
The three peaks form a straight line with a position angle of $130^\circ$.  The
inclination angle of the outflow is not well-determined.  \citet{enoch09b}
performed radiative transfer modeling of continuum emission in multiple
wavebands from $3.6\:\mu$m to 1 mm.  Their best-fit inclination angle for the
outflow is $75^\circ$, although angles in the range $65^\circ$--$80^\circ$ also
produce reasonable fits. We adopt an inclination angle of $72.5\pm7.5^\circ$. No
candidate pseudodisk has been observed for Serp-FIR1.  High-resolution
interferometric observations show a very round or unresolved core at 3 mm
\citep{williams00,hogerheijde99,enoch09b} and also at 1 mm
\citep{hogerheijde99,enoch09b}.  If the pseudodisk axis is close to the outflow
axis (see \S\,\ref{sec:outflow-pseudodisk}), then this could be a projection
effect; a flattened envelope would appear round if viewed face-on.  There are no
measurements of an infall radius for Serp-FIR1. Because the model from
\citet{enoch09b} is of a rotating, collapsing sphere, we adopt their best-fit
envelope radius of 5,000 AU.

\begin{figure}[hb]

\plotone{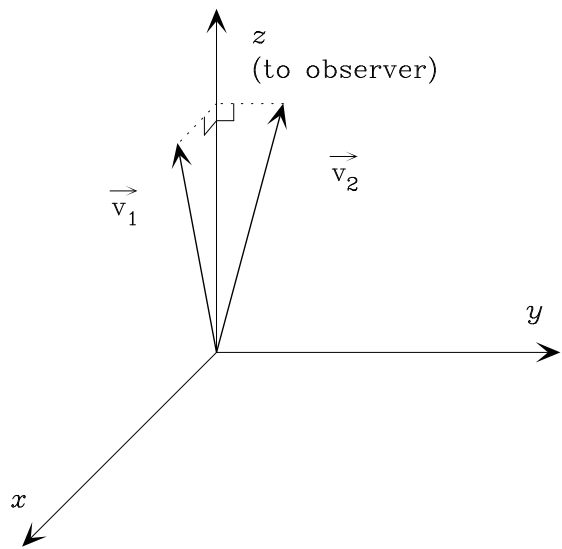}

\caption{\label{fig:cartoon2} Cartoon showing how viewing angle can alter the
projected separation between two vectors. $\vec v_1$ lies in the $xz$ plane and
$\vec v_2$ lies in the $yz$ plane.  To an observer looking along the $z$ axis,
the projected separation angle between the two vectors is $90^\circ$, even
though their true separation angle is $21^\circ$.}

\end{figure}

\begin{deluxetable*}{lcrrrc}
\tablewidth{0pt}
\tablecaption{\label{tab:mag}Source Properties Used For Combining Results}
\tablehead{\colhead{Source} & \colhead{No.\ of} & 
\colhead{Mean $B$\tablenotemark{a}} & \colhead{$\phi$\tablenotemark{b}} &
\colhead{Inclination\tablenotemark{c}} & 
\colhead{Pixel Scale\tablenotemark{d}}\\
\colhead{} & \colhead{Vectors} & \colhead{(deg)} & \colhead{(deg)} & 
\colhead{(deg)} & \colhead{(Infall Radius)}}
\startdata
B335       & 1 & $149\pm15$           & $39\pm15$           & $9\pm\phantom{1}1$  & 0.288 \\
IC348-SMM2 & 8 & $137\pm\phantom{1}5$ & $9\pm\phantom{1}5$  & $10\pm\phantom{1}5$ & 0.380 \\
L1157      & 1 & $142\pm\phantom{1}9$ & $23\pm\phantom{1}9$ & $9\pm\phantom{1}5$  & 0.363 \\
L1448-IRS2 & 4 & $147\pm\phantom{1}5$ & $12\pm\phantom{1}5$ & $33\pm\phantom{1}7$ & 0.276 \\
L1527      & 9 & $49\pm\phantom{1}4$  & $41\pm\phantom{1}4$ & $7\pm\phantom{1}1$  & 0.250 \\
L483       & 5 & $93\pm\phantom{1}6$  & $33\pm\phantom{1}6$ & $40\pm10$           & 0.238 \\
Serp-FIR1  & 3 & $62\pm\phantom{1}5$  & $68\pm\phantom{1}5$ & $72.5\pm7.5$        & 0.788
\enddata

\tablenotetext{a}{Mean magnetic field position angle derived from SHARP data in
the present paper and \paperone.  See \S\,\ref{sec:overview} for details.}

\tablenotetext{b}{Difference between the mean magnetic field position angle and
the pseudodisk apparent minor axis.  For Serp-FIR1, the outflow axis serves as a
proxy for the pseudodisk minor axis.}

\tablenotetext{c}{Inclination angle of outflow}

\tablenotetext{d}{Size of a single pixel of the final Stokes parameter maps
as a fraction of the infall radius.}
\end{deluxetable*}

Our results for Serp-FIR1 are shown in Figure \ref{fig:sharpSerp}.  Despite the
high inclination angle for the Serp-FIR1 outflow, for completeness we plot the
vectors on the \als{} model in Figure \ref{fig:sharpSerp}.  The position angle
of the pseudodisk plane is taken to be perpendicular to the outflow axis since
no pseudodisk is detected (see discussion in \S\,\ref{sec:outflow-pseudodisk}).
The magnetic field is well-ordered.  Its mean direction is $62\pm5^\circ$,
nearly perpendicular to the outflow axis, and thus also to the assumed
pseudodisk symmetry axis.  This may indicate that magnetic fields do not
regulate star formation in this core.  Alternatively, it may be a projection
effect caused by the outflow pointing nearly parallel to the line-of-sight.  To
understand this, note that the perceived angle between two vectors is heavily
dependent on viewing angle. This can be readily seen by considering the
situation depicted in Figure \ref{fig:cartoon2}.  An observer along the $z$ axis
will measure the projected separation angle between the two vectors to be
$90^\circ$, even though the true angle between them is much smaller.  We
conclude that the high inclination of the Serp-FIR1 outflow makes this source a
poor test of the basic predictions of magnetically regulated core-collapse (see
\S\,\ref{sec:introduction}).  We will further explore the effects of viewing
angle on projected separation in \S\,\ref{sec:align}.

Polarization towards Serp-FIR1 has been detected at $850\:\mu$m by
\citet{matthews09}.  In general, their polarization measurements trace the
magnetic field in the cloud rather than in the Serp-FIR1 core. All but one of
their vectors correspond to regions where the measured \threefifty{} flux is
less than $20\%$ of the peak flux.  This one vector implies a magnetic field
position angle of $63.3^\circ$, almost identical to our mean field direction.
\citet{hull13} measured 1.3 mm polarization in Serp-FIR1 on spatial scales much
smaller than those studied in the present paper.  Their mean magnetic field
direction differs by nearly $90^\circ$ from our own.  This will be discussed in
more detail in \S\,\ref{sec:discussion}.

\section{Combined Analysis of Seven Cores}\label{sec:seven}

In this section we will combine results for the four sources discussed in
\S\,\ref{sec:maps} together with results for the three sources from \paperone{}
(B335, IC348-SMM2, and L1527). We will compare results obtained from this
combined sample of seven cores observed with SHARP with the predictions of
magnetically regulated core-collapse models. We will need several additional
pieces of information to carry out these comparisons; these are given in Table
\ref{tab:mag}.  For B335, IC348-SMM2, and L1527 we computed mean magnetic field
position angles from the data in Table 1 of \paperone, using the method
described in \S\,\ref{sec:overview} of the present paper.  We obtained
corresponding pseudodisk position angles from \S\,3.1 of \paperone, and
corresponding outflow inclination angles from Table 2 of \paperone{} or from the
literature.  For B335 we used an outflow inclination $i=9\pm1^\circ$
\citep{moriarity89} and for L1527 we used $i=7\pm1^\circ$ \citep{zhou96}.  The
inclination angles for two of the sources, IC348-SMM2 and L1157, do not have
uncertainties in the literature.  For these we used the average of the
uncertainties in inclination on the other five cores, which was $5^\circ$.

\paperone{} used both SHARP \threefifty{} and SCU-POL $850\:\mu$m polarization
data when analyzing the source B335.  In the present paper we will only use the
\threefifty{} data.  With a total of seven cores, our survey now has a
sufficiently large sample for us to be able to rely exclusively on the SHARP
data, providing a homogeneous dataset.  Emission at $850\:\mu$m traces cooler
dust in comparison with \threefifty{} emission.  For example, the ratio of
\threefifty{} to $850\:\mu$m flux is three times larger for 20 K dust than it is
for 10 K dust.  Thus, the \threefifty{} polarization data preferentially trace
warmer regions closer to the protostar while the $850\:\mu$m polarization data
preferentially trace cooler regions further away from the central source.

\subsection{Correlation Between Outflow Axis and
Pseudodisk Symmetry Axis}\label{sec:outflow-pseudodisk}

We find a strong correlation between the projected outflow axis and the
projected pseudodisk symmetry axis for all cores with a measured pseudodisk
(i.e., all except Serp-FIR1).  Using data from Table \ref{tab:prop} of the
present paper and Table 2 and \S\,3.1 of \paperone, we find that the differences
in angle between these two axes are: $21^\circ$ (L483), $10^\circ$ (L1157),
$3^\circ$ (L1448-IRS2), $0^\circ$ (L1527), $17^\circ$ (IC348-SMM2), and
$20^\circ$ (B335).  The mean of these six values is $12^\circ$.  This tight
correlation gives us confidence in using the outflow axis as a proxy for the
pseudodisk symmetry axis in Serp-FIR1, as we did in \S\,\ref{sec:serp} above. In
\S\,\ref {sec:align} we will test for a correlation between the pseudodisk and
magnetic field directions.  For this purpose we will need the inclination of the
pseudodisk axis.  Because this is unknown, we will exploit the tight correlation
between pseudodisk and outflow axes by using the inclination of a given source's
outflow as a proxy for the inclination angle of that source's pseudodisk axis.
Furthermore, for Serp-FIR1 (only) we will again use the position angle of the
outflow as a proxy for the position angle of the pseudodisk axis, while for the
remaining six sources we will use our measured pseudodisk axis position angles.

\subsection{Correlation Between Mean Magnetic Field and
Pseudodisk Symmetry Axis}\label{sec:align}

We define $\phi$ to be the projected plane of sky separation angle between the
mean magnetic field and the pseudodisk symmetry axis.  The value of $\phi$ for
each core is listed in Table \ref{tab:mag}.  Since these values range from
$9^\circ$ to $68^\circ$, it appears that any correlation is much less evident
than the correlation we found between outflow and pseudodisk axis.  However, six
of the seven $\phi$ values are less than $45^\circ$, and the remaining value
corresponds to Serp-FIR1, which, as we discussed in \S\,\ref{sec:serp}, has high
outflow inclination making this source a poor test for intrinsic 3D alignment.
In the remainder of this section, we will explore quantitatively a possible
correlation between the 3D core magnetic field angle and the 3D pseudodisk
symmetry axis for our sample of seven sources.  We define $\alpha$ to be the
true 3D separation angle between magnetic field and pseudodisk symmetry axis and
$i$ to be the inclination angle of the pseudodisk symmetry axis.  (Recall that
we will use outflow inclination as a proxy for pseudodisk inclination, as
discussed in \S\,\ref{sec:outflow-pseudodisk}.)  In our analysis, we assume
that each source has the same $\alpha$, and we use our data to obtain a best
estimate for this parameter.

\begin{figure}

\plotone{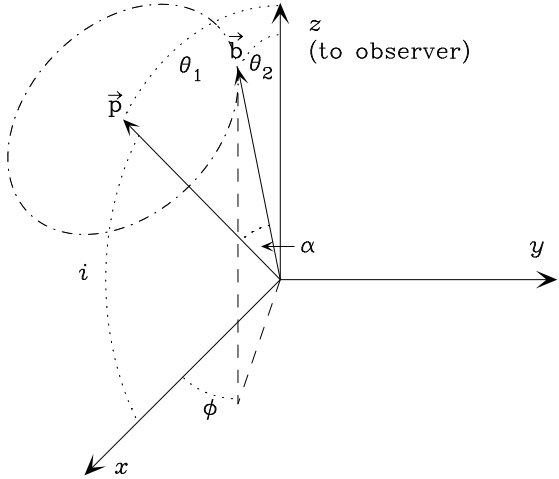}

\caption{\label{fig:cartoon} Cartoon showing the geometry for deriving the
probability density function for the projected separation angle, $\phi$, between
two vectors $\vec p$ and $\vec b$, as a function of the inclination $i$ of $\vec
p$ and the intrinsic separation angle $\alpha$ between $\vec p$ and $\vec b$.
Vector $\vec p$ represents the pseudodisk axis and is constrained to lie in the
$xz$ plane, while vector $\vec b$ represents the average magnetic field
direction.  $\alpha$ and $i$ are held constant when computing the probability
density function for $\phi$.}

\end{figure}

Consider two vectors: $\vec p$, representing the pseudodisk symmetry axis, and
$\vec b$, representing the mean magnetic field direction. Figure
\ref{fig:cartoon} shows the coordinate system and nomenclature used. Vectors
$\vec p$ and $\vec b$ can be written in $xyz$ components as:
\begin{eqnarray}
\vec p &=& p(\sin \theta_1,\:\: 0,\:\: \cos \theta_1)\\
\vec b &=& b(\sin \theta_2\cos \phi,\:\: \sin \theta_2
\sin \phi,\:\: \cos \theta_2).
\end{eqnarray}

Now consider a new coordinate system that is rotated about the $y$ axis by
$\theta_1$ degrees such that the new $z$-axis, which we call $z'$, is
aligned with $\vec p$.  Note that under this rotation $y' = y$.  In the
new $x'y'z'$ coordinate system $\vec b$ is written as:
\begin{eqnarray}
\vec b & = & b(\cos \theta_1 \sin \theta_2 \cos \phi -\sin \theta_1
\cos \theta_2,\:\: \sin \theta_2\sin \phi,\nonumber \\
       &   &  \:\: \sin \theta_1 \sin \theta_2 \cos \phi + \cos \theta_1 \cos \theta_2) \\
& \equiv & b (\sin \alpha \cos \phi',\:\: \sin \alpha \sin \phi',\:\: \cos
\alpha),
\end{eqnarray}

\noindent where in the primed coordinate system $\alpha$ takes the place of
$\theta_2$ and $\phi'$ takes the place of $\phi$.  If one holds $\alpha$ fixed
while rotating $\vec b$ about $\vec p$, then $\phi'$ varies from $0^\circ$
to $360^\circ$.  We can set the three components of $\vec
b$ equal to each other:
\begin{eqnarray}
\sin \alpha \cos \phi' & = & \cos \theta_1 \sin \theta_2 \cos \phi -
\sin \theta_1 \cos \theta_2\\
\sin \alpha \sin \phi' & = & \sin \theta_2 \sin \phi \\
\cos \alpha & = & \sin \theta_1 \sin \theta_2 \cos \phi +
\cos \theta_1 \cos \theta_2.
\end{eqnarray}

Using these three equations we can then solve for the projected separation,
$\phi$:
\begin{equation}
\label{eq:fit}
\tan \phi = \frac{\tan \alpha \sin \phi'}{\sin \theta_1 + \cos \theta_1
\tan \alpha \cos \phi'}.
\end{equation}

Therefore, by assuming an inclination angle $i$, where $\theta_1 = 90-i$, and
assuming a separation angle $\alpha$, we can compute $\phi$ as a function of
$\phi'$.  As $\phi'$ varies from $0^\circ$ to $360^\circ$, $\phi$ will take on a
range of values.  The probability density function for $\phi$ can then be
derived by making the reasonable assumption that all values of $\phi'$ are
equally likely.

\begin{figure*}

\begin{center}
\includegraphics[angle=-90,width=\textwidth]{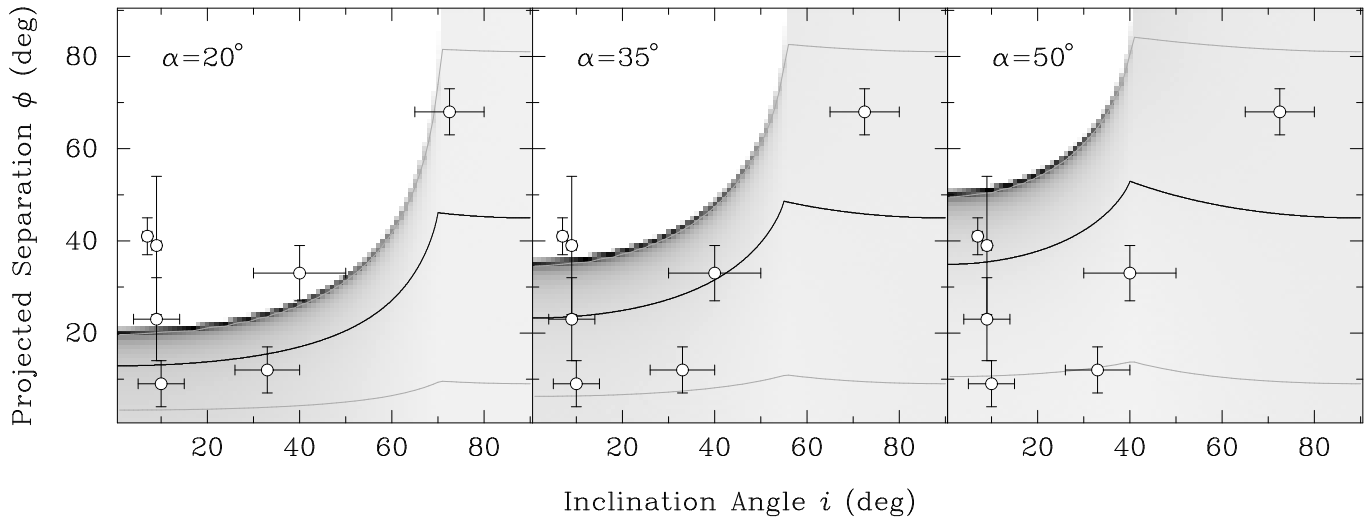}
\end{center}

\caption{\label{fig:density} Probability density functions (grayscale) of
the projected separation angle, $\phi$, between the mean magnetic field and
observed pseudodisk symmetry axis as a function of pseudodisk inclination
angle $i$.  Three intrinsic separation angles, $\alpha$, are shown. ($\alpha$
is defined as the 3D separation angle between the magnetic field and
pseudodisk axis.) The black line shows the mean separation angle as a
function of inclination.  The two gray lines correspond to $10\%$ (lower)
and $90\%$ (upper) probability (i.e., for a given inclination the specified
percentage of projected separations lie below the given curve).  The data
points represent the seven cores that have been observed by SHARP.  See
\S\,\ref{sec:align} for details.}

\end{figure*}

We computed the probability density functions (PDFs) for three values of
$\alpha$ and 901 equally spaced values of $i$ ($0^\circ$--$90^\circ$ in steps of
$0.1^\circ$).  In Figure \ref{fig:density} we show these results as three plots,
one for each value of $\alpha$.  The PDFs are shown in grayscale where dark
means most likely value of $\phi$ for each $i$.  The black curve shows
$\bar{\phi}(i,\alpha)$, the mean value of $\phi$ for each $i$ and $\alpha$
(averaged over all $\phi'$) and the lower (upper) gray curve marks the value of
$\phi$ lying above $10\%$ ($90\%$) of the integrated probability for each $i$.
The data for our seven cores, taken from Table \ref{tab:mag}, are superposed on
the probability density maps.  Recall that for each core the outflow inclination
has been used as a proxy for the unknown pseudodisk inclination $i$ (see
\S\,\ref{sec:outflow-pseudodisk}).

From an examination of Figure \ref{fig:density} it is clear that some values of
$\alpha$ fit the data better than do others.  For $\alpha=20^\circ$ some points
appear in a region forbidden by the models and are far away from the average
curve.  The opposite extreme happens for $\alpha=50^\circ$, where two of the
points lie below the $10\%$ curve.  Qualitatively, $\alpha=35^\circ$ appears
more likely since all points lie near the average curve.  Two points lie in the
forbidden region, but they are very close to the border, where the PDF is
largest.

A rigorous analysis aimed at finding the best-fit $\alpha$ by taking into
account the full density distribution as well as the uncertainties in
both projected separation $\phi$ and inclination $i$ is beyond the scope of this
paper. Instead, we will crudely estimate the best-fit $\alpha$ by
minimizing the chi-squared difference between our data and 
$\bar{\phi}(i,\alpha)$.  We created a grid of models with $\alpha$ ranging
from $0^\circ$ to $90^\circ$ in steps of $0.1^\circ$.  For each model we
computed $\chi^2$, defined as
\begin{equation}
\chi^2 = \sum_{j=1}^7 \frac{[\phi_j - \bar{\phi}(i,\alpha)]^2}{\sigma_{\phi_j}^2},
\end{equation}

\noindent where $\phi_j$ is the projected separation for an individual 
source and $\sigma_{\phi_j}$ is the uncertainty in the projected separation. 
For this simple analysis we set all the $\sigma_{\phi_j}$ to unity when 
computing $\chi^2$, thereby giving each point equal weight.  Thus, our simple 
approach amounts to a least squares fit to the average curve 
$\bar{\phi}(i,\alpha)$. The model with the smallest $\chi^2$ has $\alpha = 
35.0^\circ$, which is our initial crude estimate for the best-fit value of 
$\alpha$.

Our analysis method may lead to a bias in our best-fit value for
$\alpha$. To see this, note that it favors models in which the data points lie
close to the average curve, whereas in reality we expect the data points  for a
given $i$ to follow a distribution given by the PDF.  Our crude approach could
thus penalize high values for $\alpha$, for which the data are expected to have
a relatively larger spread from the model average curve (see Fig.\ 
\ref{fig:density}).  To explore the magnitude of this possible bias, we carried
out Monte Carlo simulations as follows: We adopted an assumed value
$\alpha_{true}$, then using the measured inclinations $i$ for our seven
sources, we computed from Equation \ref{eq:fit} seven values of the projected
separation $\phi$ by giving each source a random $\phi'$.  We then fit these
simulated data to find $\alpha_{fit}$.  We repeated this 10,000 times each for
$\alpha_{true}$ in the range $20^\circ$ to $50^\circ$ in steps of $1^\circ$.  We
then computed the mean $\alpha_{fit}$ for each input $\alpha_{true}$ and fit the
data to a straight line to obtain the following relation:
\begin{equation}
\label{eq:bias}
\alpha_{true} = 0.91\alpha_{fit} + 4.24^\circ.
\end{equation}

The mean absolute difference between $\alpha_{true}$ computed via 
Equation \ref{eq:bias} and the value input to the Monte Carlo simulations is 
only $0.26^\circ$. Starting with our initial best-fit $\alpha$ of $35.0^\circ$, 
we used Equation \ref{eq:bias} to find a bias-corrected best-fit $\alpha$ of 
$36.1^\circ$.

We are now in a position to compute the probability of obtaining a
bias-corrected best-fit $\alpha$ at or below $36.1^\circ$ by pure chance. We
consider the case where $\vec b$ points in a random direction in 3D space
compared to $\vec p$ (see Fig.\ \ref{fig:cartoon}).  Note that this is not
equivalent to choosing uniformly distributed random values for $\alpha$ and
$\phi'$ because such a distribution will sample the unit sphere non-uniformly
(the differential unit of surface area is $\sin \alpha\,
\mathrm{d}\alpha\,\mathrm{d}\phi'$).  We found that when $\alpha$ and $\phi'$
are chosen such that the unit sphere is sampled uniformly, then the resulting
distribution of $\phi$ (computed from Eq.\ \ref{eq:fit}) is also uniform,
regardless of $\theta_1$.  Therefore, generating random directions of $\vec b$
reduces to merely generating random values of $\phi$ directly.

We ran 10,000 random $\vec b$ Monte Carlo simulations.  In each
simulation, we held the inclination angles of each of our seven cores fixed at
the values shown in Table \ref{tab:mag} and Figure \ref{fig:density},
and for each core we chose a random value of the projected separation
$\phi$ ($0^\circ$ to $90^\circ$, uniform distribution).  Then we computed the
best-fit $\alpha$ for each simulation.  Only $4.21\%$ of the models
yielded a bias-corrected best-fit $\alpha$ at or below $36.1^\circ$. 
Alternatively, $95.79\%$ of models with random projected separation angles led
to a best-fit $\alpha$ greater than $36.1^\circ$.  Therefore, our analysis
indicates that we have detected a positive correlation between the core
magnetic field direction and the pseudodisk symmetry axis with $\sim$$96\%$
confidence.  Additional sources of error are considered in
\S\,\ref{sec:bias} below, leading to some modifications to the above conclusions
regarding $\alpha$ and our confidence level.

\begin{figure*}[ht]
\plotone{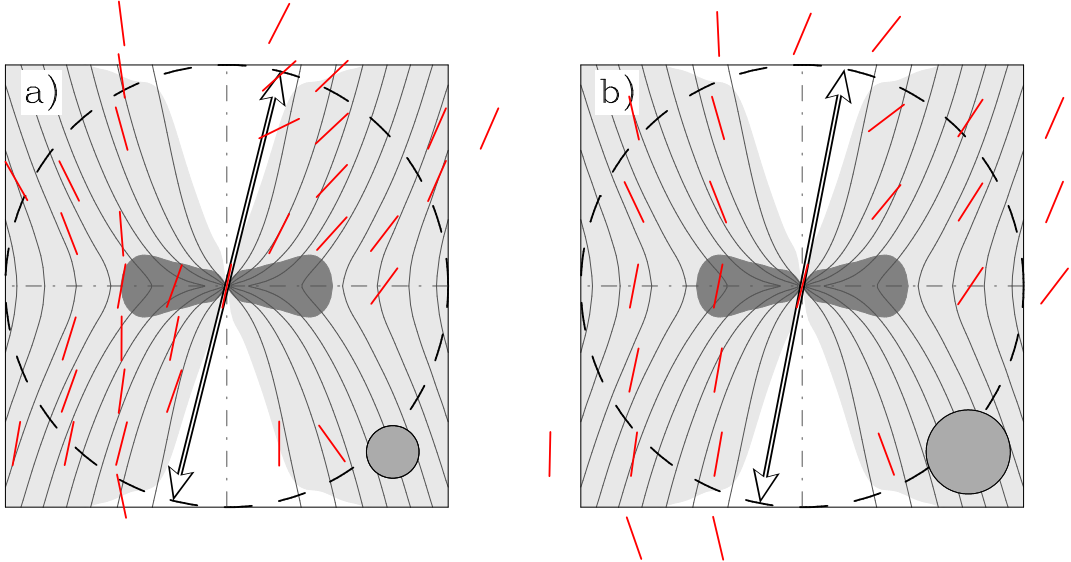}

\caption{\label{fig:allen} Source-averaged magnetic field map (red bars)
obtained by combining data from six cores (excluding Serp-FIR1), superposed on
Figure 8(c) from \citet{allen03b} showing the pseudodisk and magnetic field
lines.   Before combination, all source maps were rotated so that the pseudodisk
axis was horizontal and the blue-shifted lobe of the molecular outflow pointed
roughly upward. The gray circle in the bottom right shows the FWHM of the
Gaussian kernel used for combining data.  \emph{a)} Map obtained using a
Gaussian kernel with a FWHM of 0.238 infall radii.  The mean magnetic field
angle (white outlined vector) is $166^\circ$ (offset by $14^\circ$ from the
pseudodisk symmetry axis).  \emph{b)} Map obtained using a Gaussian kernel with
a FWHM of 0.38 infall radii.  The mean magnetic field angle is $169^\circ$
(offset by $11^\circ$ from the pseudodisk symmetry axis). See
\S\,\ref{sec:combine} for details.}

\end{figure*}

\subsection{Combining Polarization Maps to Increase
Signal-to-Noise}\label{sec:combine}

In the previous subsection we utilized the mean magnetic field direction for
each of our seven cores, as computed using $2\sigma$ polarization measurements
from Table \ref{tab:pol} of the present paper and Table 1 of \paperone. However,
our Stokes core maps hold more information than what is contained in just the
points with $2\sigma$ detections.  We can make use of this information by
combining the maps for many cores into a single source-averaged magnetic field
map.  Such a map will naturally have an enhanced signal-to-noise ratio.   Our
goal is to compare this combined map to Figure 8(c) of \als, which is the model
shown in Figures \ref{fig:sharpL483} to \ref{fig:sharpSerp}.  As discussed in
\S\,\ref{sec:serp}, the high inclination angle of Serp-FIR1 makes this core a
poor test of  magnetically regulated core-collapse, and also makes it unsuitable
for overlaying on Figure 8(c) of \als.  For these reasons, it is excluded from
the following analysis.

We start with the combined maps for Stokes $I$, $Q$, and $U$ and their
associated errors for each core (see \S\,\ref{sec:observations}).  We exclude
sky positions where the flux is less than $25\%$ of the peak, just as was done
in \paperone{} and in earlier sections of the present paper. Next, we rotate the
pixel positions in each core so that the pseudodisk lies horizontal and the
blue-shifted lobe of the molecular outflow lies in the top half of the rotated
figure.  Note that when rotating $Q$ and $U$ by an angle $\theta$, the values
become intermixed by a rotation matrix having angle $2\theta$:
\begin{equation}
\left[\begin{array}{c}
Q'\\
U'
\end{array}
\right] = \left[\begin{array}{cc}
\cos 2\theta & -\sin 2\theta \\
\sin 2\theta & \phantom{-}\cos 2\theta
\end{array}
\right] \left[\begin{array}{c}
Q \\
U
\end{array}
\right].
\end{equation}

\noindent We next compute $q=Q/I$ and $u = U/I$ and their associated errors
for each core.

Just as for Figs. \ref{fig:sharpL483}--\ref{fig:sharpSerp}, our magnetic field
maps must be scaled to account for distance and infall radius before they can be
compared with the \als{} model.  Thus, we convert the rotated pixel offsets for
each core onto the same scale by dividing the pixel scale in arcseconds
($9\farcs5$) by the core's infall radius measured in arcseconds.  The computed
pixel scale for each core is given in Table \ref{tab:mag}.  With the combined
data from all six cores we have many independent polarization measurements all
ready for superposition onto the \als{} model.  Next these measurements need to
be combined and sampled on a uniform grid to increase the signal-to-noise.

To combine all these polarization measurements into one polarization map we
overlay a grid on the data and at each grid point compute the weighted averages
$\bar q$ and $\bar u$ as well as the associated errors $\sigma_{\bar q}$ and
$\sigma_{\bar u}$.  We use a Gaussian kernel with weighting based on distance
from the grid point and the errors in the individual measurements.  The kernel
FWHM is set equal to the grid spacing and we use a cutoff radius also equal to
the grid spacing.  We have freedom in choosing the grid spacing, so we opt to
use both extremes from Table \ref{tab:mag}, yielding two different
source-averaged maps.  L483 has the smallest grid spacing (highest resolution)
while IC348-SMM2 has the largest spacing (lowest resolution).  B335, L1448-IRS2,
and L1527 have spacings similar to that of L483 while L1157's spacing is close
to that of IC348-SMM2.

After creating regularly sampled maps of $\bar q$, $\bar u$, $\sigma_{\bar q}$,
and $\sigma_{\bar u}$ we compute the polarization and magnetic field direction
using the same techniques, cutoffs, and debiasing as for the analysis described
in \S\,\ref{sec:observations}.  We obtain 35 vectors for the high resolution map
and 26 for the low resolution map.  These magnetic field maps are shown in
Figure \ref{fig:allen}.  The gray circles at the bottom right of each panel show
the Gaussian kernel used to create each map.  We also compute and plot the mean
magnetic field direction (Equal Weight Stokes Mean; see \S\,\ref{sec:overview}).
The mean field angle is $166^\circ$ for the data in Figure \ref{fig:allen}a and
$169^\circ$ for Figure \ref{fig:allen}b.

The combining of our sources to produce a source-averaged magnetic field map for
a Class 0 protostar gives us a better sampling of the characteristic magnetic
field structure than was possible for any individual source. The appearance of
Figure \ref{fig:allen} is consistent with our result from the previous
subsection; the mean magnetic field direction is nearly parallel to the
pseudodisk symmetry axis (within $15^\circ$ for both the low- and
high-resolution maps).  Furthermore, we see hints of a pinch in the field, as
predicted by magnetically regulated core-collapse models.  The pinch appears to
be stronger than predicted on the right side of each map and weaker than
predicted on the left side. Nevertheless, all four quadrants of each map show a
tendency for the field to be drawn inward as we approach the pseudodisk plane,
in qualitative agreement with model predictions.

\subsection{Correlation Between Mean Magnetic Field
and Outflow Axis}\label{sec:outflow-bfield}

Because the outflow axis is strongly correlated with the pseudodisk symmetry
axis (\S\,\ref{sec:outflow-pseudodisk}) and the pseudodisk symmetry axis is
preferentially aligned with the magnetic field (\S\,\ref{sec:align}), we
anticipate a positive correlation between outflow axis and magnetic field. We
can address this quantitatively by repeating the analyses of \S\,\ref{sec:align}
and \S\,\ref{sec:combine} using the outflow axis instead of the pseudodisk
symmetry axis.  When computing the projected separation $\phi$ using the
outflow axis, we find the bias-corrected best-fit value of $\alpha_o$ to be
$34.4^\circ$ (here denoted $\alpha_o$ to distinguish it from $\alpha$ computed
using the pseudodisk).  From Monte Carlo simulations we estimate the probability
of obtaining such a value by chance to be $2.87\%$, corresponding to a
confidence level of $\sim$$97\%$.  These results are nearly the same as the values
obtained using the pseudodisk axis, which were $\alpha=36.1^\circ$ and
$\sim$$96\%$ confidence.  We also stacked and combined the cores, as in
\S\,\ref{sec:combine}, but this time rotated each core so that the outflow axis
points straight up and down with the blue lobe of the molecular outflow still in
the top half of the rotated figure. The mean magnetic field angle is $160^\circ$
at high resolution and $165^\circ$ at low resolution.  These angles are similar
to the $166^\circ$ and $169^\circ$ values obtained when setting the observed
pseudodisk major axis to be exactly horizontal. Furthermore, the magnetic field
maps are nearly identical to those shown in Figure \ref{fig:allen}.  We conclude
that we have found evidence of a correlation  between magnetic field direction
and outflow axis, with a similar degree of confidence as for the
pseudodisk-magnetic field correlation discussed earlier.

\subsection{Sources of Error}\label{sec:bias}

One possible source of error is the slight difference in selection of vectors
between \paperone{} and the present paper.  \paperone{} considered vectors
having $p/\sigma_p \geq 2$ before debiasing to be polarization detections while
in the present paper we require vectors to pass that threshold after debiasing.
Since debiasing lowers $p/\sigma_p$, when we computed the mean magnetic field
angles for \paperone{} cores we used some vectors that would not have been
selected under our new criterion.  If we require $p/\sigma_p \geq 2$ after
debiasing for the cores in \paperone{}, we lose five vectors (1 in L1527, 3 in
IC348-SMM2, and 1 in B335).  Without these vectors, the mean magnetic field
angle changes from $49^\circ$ to $47^\circ$ in L1527 and from $137^\circ$ to
$147^\circ$ in IC348-SMM2.  There is only one high-flux vector in B335, so
applying the stricter criterion results in the loss of this source for analysis,
leaving only six sources for fitting $\alpha$.  The bias-corrected
best-fit $\alpha$ decreases to $31.5^\circ$ ($\sim$$97\%$ confidence) and the
bias-corrected best-fit $\alpha_o$ decreases to $26.2^\circ$ ($\sim$$98\%$
confidence). Thus, if we apply the stricter selection criterion to the data
from \paperone{}, the statistical significance of the correlation between mean
magnetic field direction and pseudodisk symmetry axis increases slightly.  The
same is true for the  correlation between mean field direction and outflow axis.

In \S\,\ref{sec:observations} we discussed the improved error analysis used in
the present paper.  Is it possible that this improved error analysis, if applied
to the data in \paperone, would significantly alter those results?  To test
this, we reprocessed the data for L1527 using the new method.  There are seven
polarization detections in common between the old and new methods.  The median
difference in angle for these seven polarization vector pairs is $5^\circ$,
which is less than the median uncertainty of $10^\circ$.  Furthermore, the
difference in mean magnetic field angle is less than one degree.  Therefore, it
does not appear that the change in error analysis method between \paperone{} and
the present paper significantly impacts our results.

Observational uncertainties in inclination angle and projected
separation are another source of error.  Small variations in these quantities
may lead to large changes to our best-fit $\alpha$.  To test the robustness of
our fits, we again ran Monte Carlo simulations.  In each simulation, each
source's inclination and projected separation angle were varied randomly using a
Gaussian weighting with standard deviation equal to the corresponding errors
listed in Table \ref{tab:mag}.  The resulting mean and standard deviation of the
bias-corrected best-fit values of $\alpha$ and $\alpha_o$ are $35.4^\circ \pm
3.9^\circ$ and $34.6^\circ \pm 4.5^\circ$, respectively.  Therefore, our
best-fit $\alpha$ and $\alpha_o$ are robust to within $\sim$$4^\circ$. 
Uncertainties in inclination angle may also impact our estimated confidences. 
We re-ran the Monte Carlo simulations described in the last two paragraphs of
\S\,\ref{sec:align}, this time allowing the inclination angles to vary with a
random Gaussian weighting.  The probability of obtaining a bias-corrected
best-fit $\alpha$ of less than or equal to $36.1^\circ$ by pure chance increases
to 5.25\%, and the probability for $\alpha_o$ increases to 3.92\%.  Therefore,
we downgrade our earlier confidence levels for the pseudodisk-magnetic field and
outflow-magnetic field correlations to 95\% and 96\%, respectively.

Poor sampling of a pinched magnetic field can lead to error in computing the
mean field.  For example, if in Figure \ref{fig:allen} the magnetic field  were
to be only measured in the upper left and bottom right quadrants, then the
computed mean field direction would be biased counterclockwise.  Such an error
may have occurred for L1527 and B335 (shown in Figures 5 and 7 of \paperone,
respectively).  If the field in these cores is accurately described by the
\als{} model aligned with the pseudodisk symmetry axis, then the computed mean
magnetic field direction in both cores is rotated away from the symmetry axis of
the magnetic field by a large angle.  This would lead to too-large projected
separation values, $\phi$, for these cores.  Fitting artificially high values of
$\phi$ at low inclination would lead to an overestimate of $\alpha$. Because of
this possible source of error, our best-fit values for $\alpha$ and $\alpha_o$
should be treated as rough upper limits rather than best estimates. 
Folding in the uncertainties discussed in the previous paragraph, we  estimate
an upper limit on $\alpha$ of $35.4^\circ+3.9^\circ$ or $\sim$$39^\circ$.
Similarly, our estimated upper limit on $\alpha_o$ becomes $34.6^\circ +
4.5^\circ$ or $\sim$$39^\circ$.

\section{Discussion}\label{sec:discussion}

As discussed in \S\,\ref{sec:introduction}, the first prediction of magnetically
regulated core-collapse models is that a pseudodisk exists and has its symmetry
axis aligned with the core magnetic field. In \S\,\ref{sec:align} and
\S\,\ref{sec:bias} we showed that for the cores in our sample the pseudodisk
symmetry axis does tend to align with the mean magnetic field direction ($\alpha
< 39^\circ$), and we showed that this result is unlikely to be due to chance.
If the pseudodisk symmetry axes and core magnetic fields are indeed
preferentially parallel, then we can conclude that interstellar magnetic fields
must play a significant role in the collapse of molecular cloud cores. 
Turbulence may have some effect on this process, but it cannot be strong
enough to completely overcome the tendency for organized inward motion of gas
along ordered field lines.  An important avenue for future research would be to
better constrain the pseudodisk-magnetic field misalignment angle $\alpha$, as
this angle could serve as a point of comparison between observations and
theories.

The second prediction of magnetically regulated core-collapse models is that the
magnetic field should be pinched.  Our source-averaged map (Figure
\ref{fig:allen}) shows hints of a pinch in agreement with this prediction.
Furthermore, it appears that the pinch continues outside of the infall region,
although more vectors beyond the infall radius are needed to confirm this
result.  Observations of the field outside the infall region provide a way to
discriminate among magnetically regulated star formation models.  For example,
the \als{} model has a gentle pinch outside the infall region while in some
other models the field is uniform there \citep[e.g.,][]{galli93a,galli93b}.
Previous investigations \citep{li09,ward-thompson09} have found some degree of
alignment between the magnetic fields of cloud cores (traced by submillimeter
polarization) and the magnetic fields in the surrounding regions of the cloud
(traced by optical polarization). An interesting avenue for future research
would be to map the magnetic fields in the cloud regions surrounding  each of
our seven cores, for comparison with the internal core fields mapped by SHARP.
Near-infrared polarimeters such as Mimir \citep{clemens07} and SIRPOL
\citep{kandori06} can observe polarization of background stars viewed through
cloud regions surrounding a core.  Furthermore, because they operate at near-IR
wavelengths (where the extinction is lower than at optical wavelengths), they
can probe denser regions nearer to the cores which generally cannot be studied
via optical polarimetry.

In \S\,\ref{sec:outflow-bfield} we found that the axes of bipolar outflows are
preferentially aligned parallel to core-scale magnetic fields with a rough upper
limit on the characteristic misalignment angle, $\alpha_o$, of
$\sim$$39^\circ$.  Since the outflow is believed to run parallel to the axis of
the Keplerian circumstellar disk (\S\,\ref{sec:introduction}), our results
suggest a preferential alignment between the circumstellar disk rotation axis
and the core magnetic field.  As discussed in \S\,\ref{sec:introduction}, 
\citet{joos12} have argued that circumstellar disks cannot form unless their
axes are misaligned with the core magnetic fields, so it is interesting that our
results suggest an upper limit of $\sim$$39^\circ$ on this misalignment.
However, it is important to remember that the outflow, pseudodisk, and magnetic
field observations discussed in this paper pertain to structures having size
scales well above the very small scales ($\sim$few AU) where outflows are
believed to be launched. Observations at much higher resolution would seem to be
required before we can confidently constrain the alignment of the young
circumstellar disks that are presumed to be growing inside Class 0 cores.

\citet{hull13} measured 1.3 mm polarization on $\sim$1000 AU scales for a sample
of 17 low-mass protostellar cores. (We referred to this work in
\S\,\ref{sec:maps} above.)  Four of these cores are also included in our own
survey: L1157, L1448-IRS2, L1527, and Serp-FIR1 (identified as Ser-emb 6 by
\citealt{hull13}).  \citet{hull13} find mean magnetic field position angles of
$139\pm9^\circ$ in L1157 and $146\pm4^\circ$ in L1448-IRS2. These values are
consistent with our respective SHARP (larger-scale) magnetic field measurements,
within the errorbars (see Table \ref{tab:mag}).  For L1527, their mean magnetic
field is $174\pm8^\circ$, suggesting a magnetic field which is toroidal, in
contrast to the poloidal field geometry claimed in \paperone.  This may indicate
that the field in L1527 has been wrapped up azimuthally on small scales by core
rotation, a possibility discussed by \citet{hull13}.  In the last source,
Serp-FIR1, \citet{hull13} measure a mean magnetic field position angle of
$157\pm3^\circ$, nearly perpendicular to our measured value.  Since the field
measured by \citet{hull13} is nearly parallel to the outflow axis (and thus
parallel to the presumed rotation axis), azimuthal fields cannot be invoked to
explain the difference between their result and ours.  As discussed in
\S\,\ref{sec:serp}, however, the high inclination angle of Serp-FIR1 makes it a
poor test of magnetically regulated core-collapse models.  Therefore,
results for the four sources in common between \citet{hull13} and the present
paper are consistent with a picture of magnetic regulation from the large (core)
to the small ($\sim$1000 AU) scales, provided that we allow for a transition to
toroidal fields on small scales in L1527.

For their full sample of 17 cores, \citet{hull13} found no correlation between
the mean magnetic field directions and the protostellar outflow axes.  Assuming
that the outflows run parallel to the axes of the circumstellar disks, then the
\citet{hull13} results imply that the disk rotation axes are not aligned with
the $\sim$1000 AU scale magnetic fields. Their results differ from our own,
since we do find a correlation between magnetic field direction and outflow
axis. We consider three possibilities to resolve this apparent discrepancy.
First, if some of the 17 cores have toroidal small-scale fields while others are
poloidal, then this would lead to a poor correlation between outflow axis and
field direction.  \citet{hull13} consider this explanation to be insufficient,
since it would lead to a bimodal distribution for the projected angle between
outflow and magnetic field, whereas the observed distribution is mostly
consistent with a random distribution rather than a bimodal one.  A second
possibility, also discussed by \citet{hull13}, is that their sample includes
multiples which present a more complex situation than what has been modeled
(e.g., by \als), and thus may obscure any correlation between outflow axis and
field direction.  Lastly, it is important to remember that the $\sim$1000 AU
size scales mapped by \citet{hull13} are considerably smaller than the
$\sim$10,000 AU scales mapped in the present paper.  It is possible that the
magnetic field may be ordered on large scales but scrambled on smaller scales,
due to some combination of rotational, pinching, and outflow motions.  More work
is needed to understand the apparent discrepancy between the results of
\citet{hull13} and our own.

\section{Summary}\label{sec:conclusions}

We presented \threefifty{} polarization maps of four low-mass cores with Class 0
protostars: L483, L1157, L1448-IRS2, and Serp-FIR1.  We created a larger sample
by combining these results with \paperone{} results for three similar cores:
B335, L1527, and IC348-SMM2.  With this sample we were able to test magnetically
regulated models of core-collapse using sources most directly comparable with
the models; i.e.\ isolated, single, young low-mass cores with outflow axes lying
close to the plane of the sky.  This last point is very important because
projection effects can come into play when comparing sky plane components of 3D
axes.  Six of the sources have their outflow axes oriented closer to the plane
of the sky than to the line of sight, while one, Serp-FIR1, has its outflow
almost along the line of sight.  The results from our sample of seven cores are
as follows:

\begin{enumerate}

\item In \S\,\ref{sec:outflow-pseudodisk} we showed that for the six sources 
with identified pseudodisks, the mean difference in position angle between the 
outflow axis and the pseudodisk symmetry axis (i.e., pseudodisk apparent minor 
axis) is $12^\circ$.  This gives us confidence in using the outflow
inclination  angle as a proxy for the pseudodisk inclination angle.
Furthermore, for  Serp-FIR1, which has no apparent pseudodisk axis, we used the
position angle of  the outflow as a proxy for the position angle of the
pseudodisk symmetry axis.  In \S\,\ref {sec:align} we used our polarization
data, the observed position  angles of the pseudodisk symmetry axes for six of
our sources, and the above  proxies, to test for a correlation in 3D space
between pseudodisk symmetry axis  and core magnetic field. Using a
simple least-squares analysis, we estimated  the 3D separation angle $\alpha$
between these two quantities to be $36^\circ$.  Our estimate for the
probability of obtaining an $\alpha$ less than or equal to  $36^\circ$ by pure
chance is about $4\%$.  In \S\,\ref{sec:bias} we modified  these conclusions
after addressing additional sources of error.  The revised  probability for our
correlation being due to chance is 5\%.  Our revised  constraint for $\alpha$
is a rough upper limit of $39^\circ$.

\item In \S\,\ref{sec:combine} we combined polarization data for six of the
seven sources into a source-averaged magnetic field map.  We excluded Serp-FIR1
because its high inclination angle makes it a poor test of the basic predictions
of magnetically regulated models. Both the low- and high-resolution versions of
our source-averaged map have many more polarization detections than any of the
individual maps.  The mean magnetic field direction in each of the two
source-averaged maps is closely aligned (within $15^\circ$) with the pseudodisk
symmetry axis.  This is consistent with our claimed correlation between the
magnetic and pseudodisk symmetry axes. The magnetic field in the source-averaged
maps shows hints of a pinch, as predicted by the magnetically regulated models.

\item In \S\,\ref{sec:outflow-bfield}, using techniques similar to those 
summarized in point 1 above, we found a correlation between the outflow axis 
and the core magnetic field direction.  As discussed in  \S\,\ref{sec:bias}, our
crude estimate for the confidence level is $\sim$96\%  and our rough upper limit
on the misalignment angle $\alpha_o$ is  $\sim$$39^\circ$.

\end{enumerate}%

If our claimed detection of a positive correlation between core magnetic field 
direction and pseudodisk symmetry axis is correct, then magnetic fields must be 
strong enough to direct gas infall even in the presence of turbulence. 
Stronger turbulence should lead to higher values for the misalignment 
angle $\alpha$, for which we have set the rough upper limit of $39^\circ$.  
Our claimed detection of a positive correlation between core magnetic field 
direction and bipolar outflow axis might constrain theories for the formation 
of circumstellar disks, but higher resolution observations are probably 
required to observationally characterize this.

\acknowledgements

We thank the anonymous referee for providing useful feedback that 
improved the statistical analysis in this paper. This material is based upon
work at the Caltech Submillimeter Observatory, which is operated by the
California Institute of Technology under cooperative agreement with the National
Science Foundation (AST-0838261).  This research was carried out in part at the
Jet Propulsion Laboratory, which is operated by the California Institute of
Technology under contract with NASA.  We are grateful to the National Science
Foundation for supporting the operation of SHARP via grant AST-0909030.  N.L.C.\
is also supported under this grant.



\end{document}